\documentclass[prd,twocolumn]{revtex4}
\usepackage{graphicx, epsfig}
\usepackage{color}
\usepackage{mathrsfs}

\newcommand{\be}{\begin{equation}}
\newcommand{\ee}{\end{equation}}
\newcommand{\bea}{\begin{eqnarray}}
\newcommand{\eea}{\end{eqnarray}}

\newcommand{\gapp}{\mathrel{\raise.3ex\hbox{$>$}\mkern-14mu
              \lower0.6ex\hbox{$\sim$}}}
\newcommand{\lapp}{\mathrel{\raise.3ex\hbox{$<$}\mkern-14mu
              \lower0.6ex\hbox{$\sim$}}}
\newcommand{\scri}{\mathscr{I}}

\begin{document}
\title{Black Stars and Gamma Ray Bursts}
\author{Tanmay Vachaspati}
\affiliation{CERCA, Department of Physics, 
Case Western Reserve University, Cleveland, OH~~44106-7079}

\begin{abstract}
\noindent
Stars that are collapsing toward forming a black hole but are frozen
near the Schwarzschild horizon are termed ``black stars''. Collisions 
of black stars, in contrast to black hole collisions, may be sources 
of gamma ray bursts, whose basic parameters are estimated quite
simply and are found to be consistent with observed gamma ray bursts. 
Black star gamma ray bursts should be preceded by gravitational 
wave emission similar to that from the coalescence of black holes.
\end{abstract}

\maketitle


It is well-known that, from an asymptotic observer's viewpoint,
a collapsing body is forever suspended just above its Schwarzschild 
radius.  This picture may change with the inclusion of quantum 
radiation from the collapsing body as has been discussed from
many viewpoints 
\cite{BirrellandDavies, Vachaspati:2006ki,Vachaspati:2007hr}.
However, quantum effects change the picture on time scales given by 
the black hole evaporation time scale. For astrophysical bodies such 
as the sun, the evaporation time is $\sim 10^{66}$ years, which is
$\sim 10^{56}$ times the present Hubble time. Hence for astrophysical
purposes, we can ignore evaporative processes altogether and 
work within classical general relativity. Since any radiation 
from the collapsing body is redshifted by a large amount, the body 
will appear as a dark compact object. The object will appear black but 
will not be a black hole. We will call such an object a ``black star''. 
In contrast, a ``black hole'' is a vacuum solution of Einstein's
equations and there is no matter distribution inside it except
for the singularity at the origin.

When we watch for signatures from the collapse of astrophysical 
bodies, we take the asymptotic observer's viewpoint, and hence,
gravitational collapse always leads to black stars. 
In this brief note I point out that collisions of black stars 
can be a source of gamma ray bursts, and that such bursts are 
preceded by gravitational wave emission whose characteristics
should be similar to those of black hole mergers. Even though
the basic estimates for gamma ray bursts originating in black star 
collisions agree quite well with those for observed gamma ray bursts, 
this note, at least in its current form, can only be taken as a 
suggestion for pursuing this idea further. Observed gamma ray 
bursts have very complex features, and occur in many different 
sub-classes, each possibly having a different underlying origin. 
For details about gamma ray bursts, the reader is referred to the 
literature e.g. \cite{Piran:1999kx,Lamb:2000rc,Piran:2004ba}. 

To distinguish between black stars and black holes, consider
what happens when two black objects collide. If we receive photons 
from the collision, then the colliding objects were black stars because 
the empty spacetimes of two black holes would only create gravitational 
waves. If, however, we receive only gravitational waves from the
colliding objects then they can either be primordial black holes 
or black stars that have not yet collided. Therefore, at least when
black stars do collide, they can be distinguished from black holes 
by the nature of the radiation. The infall of matter on to a black 
star will also lead to electromagnetic emission due to collisions
of the matter with the matter making up the black star; matter that 
is falling into a black hole will not emit electromagnetically except
due to collision with other infalling matter, as in an accretion
disk.

\begin{figure}
  \includegraphics[height=0.35\textwidth,angle=0]{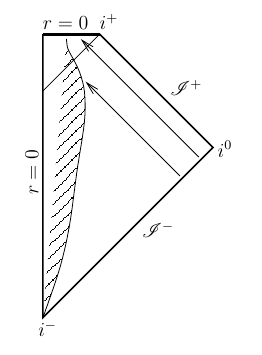}
\caption{Spacetime of a classical collapsing object (shaded region)
that appears as a black star to an observer located near $i^+$. 
Two null rays from $\scri^-$ are shown, one that collides with 
the black star and the other that does not. Similarly, an 
approaching second black star may or may not collide with 
the first black star depending on initial conditions.
}
\label{bh_spacetime}
\end{figure}

We now estimate the energy radiated from the collision of two black 
stars each of mass $M$. The radius of each of the black stars is 
$R=2GM$ and so the density is 
\begin{equation}
\rho = \frac{1}{8G^3M^2} \sim 10^{17} 
         \left ( \frac{M_\circ}{M} \right )^2
              {\rm gms / cm^3} 
              \label{bsdensity}
\end{equation}
Therefore, if two black stars of mass $M$ collide, it involves the
collision of matter at high density, and will lead to radiation 
of photons and other light particles. Since the gravitational
binding between all constituents in this system is very strong, 
we treat the collision as being totally inelastic. Then the initial 
kinetic energy gets converted to radiation resulting in the release 
of total energy
\begin{equation}
E \sim M v^2 = 10^{48} 
               \left ( \frac{M}{M_\circ} \right )
               \left ( \frac{v}{300 {\rm km/sec}} \right ) ^2
                {\rm ergs}
\label{ke}
\end{equation}
Assuming that this energy is released in a light crossing
time $\sim R/c = 2GM/c \sim 10^{-5} {\rm sec}$, which is the
only relevant length scale in the problem, the emitted
power is
\begin{equation}
P \sim 10^{53} 
         \left ( \frac{v}{300 {\rm km/sec}} \right ) ^2
            {\rm ergs/sec}
\label{power}
\end{equation}
Note that the power is independent of the mass $M$. Also, since 
the gravitational coupling is much weaker than electromagnetic 
interactions, almost all of the power will be in the form of 
photons. We can estimate the frequency of the photons by once again
treating the collision as totally inelastic. Then every proton in
the black star gets stopped on collision and the emitted photon 
energy is simply the initial kinetic energy of the proton
\begin{equation}
E_\gamma \sim m_p v^2 = 1 
           \left ( \frac{v}{300 {\rm km/s}} \right )^2 {\rm keV} 
\label{Egamma}
\end{equation}

Two black stars that approach each other with initial velocity,
$v$, may or may not collide, depending on the precise initial
conditions. This is most clearly seen on the spacetime diagram
for a collapsing star (see Fig.~\ref{bh_spacetime}). If the 
initial conditions are favorable, there will be frequent collisions; 
otherwise collisions will be rare. With our current knowledge, 
it is not possible to reliably estimate the frequency with which 
black star collisions occur. 

Even though we cannot estimate the frequency of black star
collisions, we do know that the frequency is smaller for smaller
initial velocity since then the initial conditions are not
suitable for the stars to collide. Also, we expect that the 
number of black stars falls off with higher velocity. These two 
arguments suggest that there should be a velocity at which black 
star collisions peak. In terms of gamma ray bursts, it implies
that the gamma ray bursts should have a typical photon energy.
Further, the total power emitted should scale with this photon 
energy as seen by dividing Eq.~(\ref{power}) by (\ref{Egamma}),
\begin{equation}
\frac{P}{E_\gamma} \approx 10^{62} ~ {\rm sec^{-1}}
\end{equation}
This formula does not depend on the mass of the colliding
black stars and neither on their velocities, and hence is
an invariant of the model. In a more realistic setting,
however, black star collisions will, in general, not be head-on 
and may be affected by the surrounding environment. Also, the 
emitted radiation is not spherically distributed because of
the lack of spherical symmetry in the initial system
and geometrical factors are needed to relate the emitted 
power to what an observer would see.

So far we have considered the collision of two black stars.
Similar signatures may be expected from the collision
of a black star and a normal star. 

Next consider the collision of two black objects, each of
mass $M$, where the matter does not collide but only the
spacetimes collide. This can happen with black stars for
which the initial conditions correspond to the later null
ray in Fig.~\ref{bh_spacetime}, or with primordial black holes
for which collapsing matter is not present. In this
case,  all the initial kinetic energy -- all $10^{48} {\rm ergs}$ 
(Eq.~(\ref{ke})) of it -- will be released in gravitational 
radiation, as the two spacetimes merge. However, the final
coalescence takes an infinite time and will never be seen.
Then we can never be sure if the gravitational signatures
are from colliding primordial black holes or from black stars 
whose matter is yet to collide.

If an observed gamma ray burst is indeed due to colliding 
black stars, the burst should be preceded by gravitational 
wave radiation from the coalescing spacetimes of the black 
stars.  The gravitational wave emission should be very similar 
to that calculated numerically for black hole collisions 
({\it e.g.} \cite{Baker:2006yw}), 
and the final gravitational wave burst due to coalescence 
should be replaced by the gamma ray burst when the material
of the black stars coalesce. Since the coalescence occurs
in the strong gravitational field of the black stars, we
expect electromagnetic ringing to accompany the gravitational 
ringing associated with the coalescing spacetimes. The 
characteristics of the emitted radiation will depend on the 
normal modes of the two black star system. Indeed, characteristics 
of the gravitational radiation preceding the gamma ray burst, 
together with the gamma ray burst, may allow us to infer the 
parameters of the colliding black stars and the initial conditions.

\begin{acknowledgments}
I am grateful to Martin Rees for helpful critique, and to Francesc 
Ferrer, Pablo Laguna and Irit Maor for remarks. This work was supported 
by the U.S.  Department of Energy and NASA at Case Western Reserve 
University.
\end{acknowledgments}

\section{Addendum (03/2016)}
\label{addendum}

Recently LIGO has announced the discovery of gravitational waves that match the 
signal expected from the merger of two black holes, each of mass 
$\sim 30{\rm M}_\odot$ \cite{Abbott:2016blz}. Subsequently Fermi has 
announced a gamma ray burst that is consistent with the interpretation 
of being an electromagnetic counterpart of the LIGO event with false alarm 
probability 0.0022 \cite{Connaughton:2016umz}. The Fermi event has a luminosity
in hard X-rays ($1~{\rm keV}-10~{\rm MeV}$) of $\sim 10^{49}~{\rm ergs/s}$
over a time period of $\sim 1~{\rm s}$ and following about $0.4~{\rm s}$ 
after the LIGO event. 

The black star scenario qualitatively fits the LIGO+Fermi observations.
More accurate predictions and more observational data can be expected in the near
future.

\begin{acknowledgments}
This work was supported by the U.S. Department of Energy, Office of High Energy Physics, 
under Award No. DE-SC0013605 at ASU.
\end{acknowledgments}

\end{document}